# Extending CPU-less parallel execution of lambda calculus in digital logic with lists and arithmetic

Harry Fitchett, Jasmine Ritchie, Charles Fox
School of Engineering and Physical Science,
University of Lincoln, UK

## Abstract

Computer architecture is searching for new ways to make use of increasingly available digital logic without the serial bottlenecks of CPU-based design. Recent work has demonstrated a fully CPU-less approach to executing functional programs, by exploiting their inherent parallelisability to compile them directly into parallel digital logic. This work uses $\lambda$-calculus as a hyper simple functional language to prove the concept, but is impractical for real-world programming due to the well-known inefficiencies of pure $\lambda$-calculus. It is common in language design to extend basic $\lambda$-calculus with additional primitives to short-cut common tasks such as arithmetic and lists. In this work, we build upon our previous research to examine how such extensions may be applied to CPU-less functional execution in digital logic, with the objective of advancing the approach toward practical implementation. We present a set of structures and algorithms for representing new primitives, describe a systematic process for selecting, implementing, and evaluating them, and demonstrate substantial reductions in execution time and node usage. These improvements are implemented in an open-source system, which is shown to correctly evaluate a range of representative $\lambda$ expressions.

## 1 Introduction

While clock speeds have largely plateaued, fabricatable transistor densities have continued to increase, leading to the 'new golden age' [1] of computer architecture, in which emphasis has shifted to exploiting massive parallelism in new ways. This shift brings both opportunities and challenges: we now have the hardware capability to perform many operations simultaneously, but need to rethink fundamental programming models and execution strategies to take full advantage of them.

This is especially important in emerging application domains such as robotics and the Internet of Things (IoT), where low-latency responses and edge computing are critical, but power, costs and physical size are limited. In these settings, devices often need to perform complex computations such as as digital signal processing of video and audio locally, without relying on remote servers, due to latency constraints or unreliable connectivity. In such systems, efficient, parallel execution models that can run on lightweight, embedded hardware are essential for enabling real-time autonmous decision-making, or for pre-processing and data compression to relay to external systems [2].

At the other end of the network, data center and cloud applications also stand to benefit from improved approaches to parallel computation. While GPUs and TPUs have become dominant platforms for handling large-scale workloads — such as neural networks[3, 4] and physical simulations — they require specialized programming models and are not always the best fit for general-purpose or dynamic workloads. A more flexible, general model of parallel computation could offer a viable alternative for these environments, including further processing of data streams received from robotics and IoT data.

Signal processing and control programming in particular is often performed by engineering specialists, who are focussed on mathematical concepts. Rather than requiring these programmers to learn new parallel programming paradigms, an alternative approach is to create parallel compilers from more familiar languages.

Functional languages have long been proposed as a natural way to express parallel computations [5], as their lack of mutable state and side effects makes it easier to evaluate expressions in any order. Functional programming offers a promising foundation for systems that automatically extract parallelism from conventional-looking code [6][7]. Many functional languages choose to represent programs as tree-like graphs. Computing with these graphs is referred to as graph simplification or reduction, which can be parallelized by allowing separate branches to reduce simultaneously [6][7].

$\lambda$-calculus [5] is the simplest possible functional programming language. It was proposed by Alonzo Church in 1936 as the first example and still-current definition of a universal computation system [8], meaning that any computable function can be expressed in it. A function is represented by an expression of the form $\lambda x.E$, where $\lambda x$ indicates that $x$ is a bound variable and $E$ is an expression that defines the function's body. Function application is written as $(M\ N)$, where $M$ and $N$ are $\lambda$ terms (functions or variables). The primary operation in $\lambda$-calculus is *beta reduction*, where an application $(\lambda x.E)\ N$ reduces to $E[x := N]$, meaning that the function $\lambda x.E$ is applied to the argument $N$, substituting $N$ for $x$ in $E$.

Harry Fitchett, Jasmine Ritchie, Charles Fox
School of Engineering and Physical Science,
University of Lincoln, UK

Functional languages based on $\lambda$-calculus such as Lisp, Clean, ML, and Haskell have implementations on current CPU-based hardware [9][10]. While modern CPUs include some parallel elements which can be used in execution, [11], individual CPU cores still require imperative machine code, so functional languages typically compile programs to such code, even for performing parallel graph reduction [12][13]. Term graph rewriting uses refined logical steps in lambda calculus reduction in order to speed up a costly portion of graph reduction, rewriting [14]. Techniques such as Half Combustion – a type of lazy execution [15][16] – have been used to optimize compilers. PELCR, a $\lambda$-based language compiler, uses these techniques to achieve a near 80% speedup compared to prior established methods [17][18][16]. Alternate, more abstract, expression types have been proposed to allow for predicting futures appearing in $\lambda^{as}$ these expressions can evaluate branches below them to skip unnecessary steps [18][16][19]. Other language implementations aim to reduce the amount of memory required to represent a tree, by adding expressions to represent abstract data types such as lists.

A recent study [20] introduces a CPU-less computational execution model for $\lambda$-calculus, which compiles it directly to parallel digital logic rather than using machine code or a serial CPU. This model distributes the reduction of $\lambda$-calculus programs across a reconfigurable graph of many small, parallel digital logic blocks called nodes, laid out in silicon with reconfigurable contents and connectivity somewhat like like FPGA logic blocks. In this framework, the programmer writes programs in pure $\lambda$-calculus. The programs are parsed by software into trees of expressions, then hardware nodes are configured and connected, each representing one expression. Nodes then communicate by message passing, to reduce the $\lambda$ expressions. Nodes can represent different types of expression: names, functions, or applications, with each type reacting differently to received messages. While a single node cannot compute anything alone, when connected, the collective cluster of nodes is able to reduce entire programs.

Nodes are capable of forming up to three connections with other nodes. These connected nodes are referred to as a nodes Parent, Left Child and Right Child. Nodes communicate with their assigned parents and children through a series of shared buses, exchanging two main types of information: expression and instruction data. Expression data describes a nodes current state relaying information from its internal registers and flags. Nodes store five values across two flags and three registers. The two flags include the Resolve Flag which when raised indicates no further computation is required on that nodes branch and the Irreducible Flag which when raised indicates that a nodes branch is dependent on a non-existent branch. Likewise each node has three registers: an expression register, which stores a binary key identifying the expression type, and two child pointers, which hold unique node IDs of its left and right children. By contrast nodes will exchange instruction data via the instruction bus. Unlike a CPU, instruction in this context do not refer to a series of commands a programmer has access to. Instead, to coordinate transformation and reduction steps, instructions are passed between nodes. Allowing a singular node to query values not stored within itself. Instruction data is comprised of a binary key that correlates to an operation and a target Unique Node ID which indicates the desired node.

The previous study [20] proved the basic concept of compiling and executing functional programs in digital logic. However, pure $\lambda$-calculus, while lightweight and easy to implement when compared to other functional languages, is not an ideal language for practical programming, and the study concluded that both usability and performance could be improved by expanding the range of expression types including lists and arithmetic. We here design, implement, and test extensions to the previous architecture to provide these features.

## 2 List primitives

We aim for style consistency with existing Church-encoded data structures, such as Church numerals. We begin by considering the simplest possible list, then build toward more complex forms. In most programming languages, a list grows as items are added; therefore, the simplest conceivable list is one that contains no items. To represent lists in a string-based, $\lambda$-calculus-inspired notation, we enclose their contents in brackets prefixed by a $\gamma$ symbol. Using this convention, the expression $(\gamma^0\emptyset.\emptyset)$ represents an empty list.

Once an empty list is established, we can define a list containing a single item. For example, the expression $(\gamma^0 a.\emptyset)$ represents a list whose only element is 'a'. The two expressions above scale naturally from zero to one item; however, extending this pattern to multiple items requires careful consideration. The most obvious solution appears to be simply storing the second element, say 'b', in place of the trailing NULL value. But this leads to the immediate question of where subsequent items (a third, fourth, and so on) would be placed. Instead, we follow the common structure of Church encoding, which builds composite data structures by nesting simpler structures within themselves. We represent a list containing two items, 'a' and 'b' by first representing a list containing one item storing the 'a', then nesting it within a second list which contains the 'b'. A list containing 'a' followed by 'b' is therefore written as: $(\gamma^1 b.(\gamma^0 a.\emptyset))$ To make the recursive structure more transparent, each $\gamma$ is annotated with an index indicating the position of the item it introduces, the first item has index 0, the second index 1, and so on. The following illustrates lists with one to five items:

$$
\begin{aligned}
1Item(s) &= (\gamma^0 a.\emptyset) \\
2Item(s) &= (\gamma^1 b.(\gamma^0 a.\emptyset)) \\
3Item(s) &= (\gamma^2 c.(\gamma^1 b.(\gamma^0 a.\emptyset))) \\
4Item(s) &= (\gamma^3 d.(\gamma^2 c.(\gamma^1 b.(\gamma^0 a.\emptyset)))) \\
5Item(s) &= (\gamma^4 e.(\gamma^3 d.(\gamma^2 c.(\gamma^1 b.(\gamma^0 a.\emptyset)))))
\end{aligned}
$$

Following this pattern, a formal grammar for lists in our $\lambda$-calculus-inspired system is:

$$
\begin{aligned}
\langle expression\rangle &::= \langle name\rangle \mid \langle function\rangle \mid \langle application\rangle \mid \langle list\rangle \\
\langle list\rangle &::= (\gamma\langle expression\rangle.\langle list \mid NULL\rangle)
\end{aligned}
$$

In all prior examples, list items have been names. However this definition allows list items to contain any expression potentially meaning lists can potentially house entirely programs.

Next, we must then determine how lists interact with other Church-style expressions during beta reduction. In conventional programming, accessing a list returns the element at a specified index. To support a similar mechanism in our system, we introduce two measurable properties of list expressions: depth and activity.

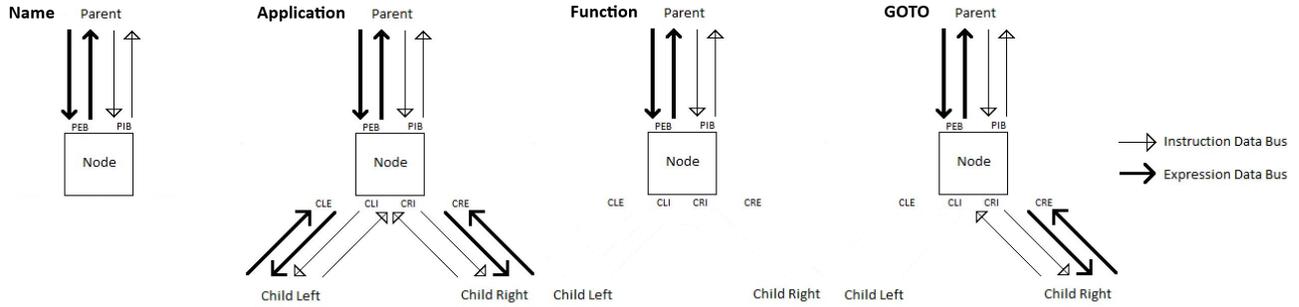

Figure 1: Existing Node Connectivity Example

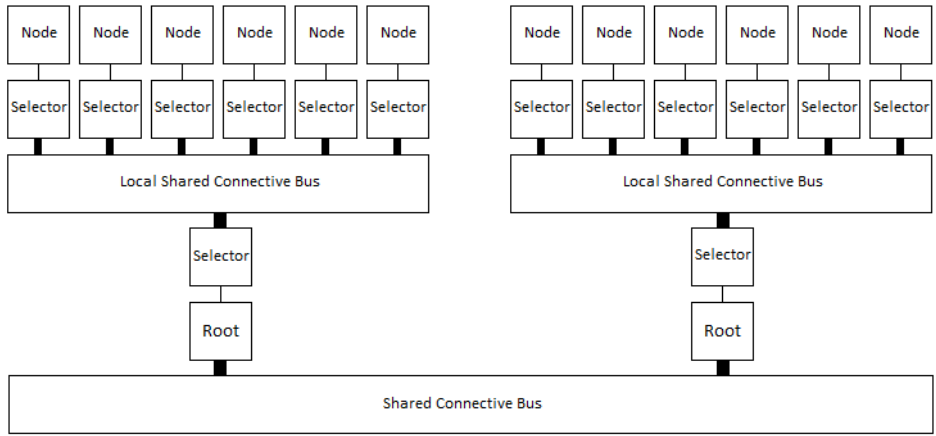

Figure 2: Cluster Connection Example

The depth of a list expression is the number of list expressions nested within it. For example, in the list $(\gamma^1 b.(\gamma^0 a.NULL))$, the inner list $(\gamma^0 a.NULL))$ contains no further lists so has depth 0. The outer list contains this inner list as its tail, giving it depth 1. This notion of depth effectively parallels the index of each element so allows us to query items in the list.

Activity, by contrast, is defined externally, before reduction begins. Activity is a Boolean property applied to list expressions. If a list expression is inactive, its item content is ignored during reduction. If a list expression is active, then during beta reduction the entire list structure collapses to the item contained at that active position. By definition, only one list expression within a list may be active at a time. We refer to the depth of this active list expression as the list structure's *activated depth*.

First a list expressions depth can be defined as the number of list expressions nested within itself. Using this definition, the lists within the following expression $(\gamma^1 b.(\gamma^0 a.NULL))$ calculate their depths as follows. The inner list contains no other lists and therefore has a depth of 0. The outer list contains a single list expression and therefore has a depth of 1. Node depth acts as an equivalent to item index, and will enable querying of the list structure.

The list $(\gamma^4 e.(\gamma^3 d.(\gamma^2 c.(\gamma^1 b.(\gamma^0 a.\emptyset)))))$ thus reduces differently – to $a$, $b$, or $d$ for chosen activated depths 0, 1 and 3 respectively.

## 2.1 Implementation

Lists are constructed from nested chains of list expressions. Each list expression points to a distinct item, and at any given moment only one item is active, that is, unsuspended within the larger graph structure. The list expression that points to this connected subgraph is called the active list, and its activity state must be stored somewhere. Rather than introduce a new register, we re-purpose a register unused by this expression type: the Resolve Flag.

The proposed solution also requires computing and storing a depth value. This value cannot be derived or maintained using the existing implementation. To support list, each node must have access to a localized adder sized to match the number of bits used to store Node IDs. With this adder, nodes can compute their parent's depth by incrementing their own depth and passing the result upward. Nodes determine their own depth by receiving it from their child; because a value of zero represents a NULL pointer in this system, the tail of the list interprets an absent child as providing a depth of zero.

Fig. 3 presents an abstract visual representation of the default output-passing behavior for list expressions. As defined in Section 2, the left child may be any expression, it can point to a single item or

Harry Fitchett, Jasmine Ritchie, Charles Fox
School of Engineering and Physical Science,
University of Lincoln, UK

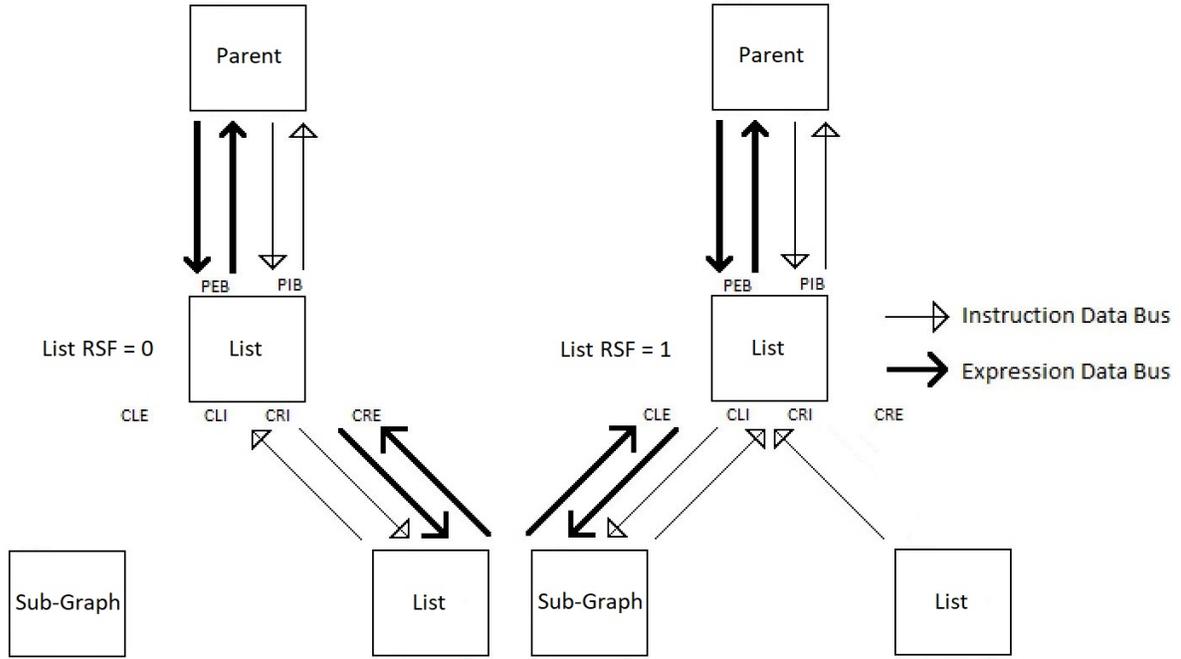

Figure 3: Outputting of Active and Inactive Lists

the root of an entire subgraph, while the right child points either to another list expression or a NULL pointer. Depending on the state of the list's Resolve Flag (its activity), the list expression passes values between its parent and either its left or right child. Unlike other expression types, list expressions do not propagate their internal status flags (Resolve and Irreducible) to their immediate neighbors. Instead, they forward the status of the nodes to which they are connected along the same paths used for value transmission. This behavior effectively links the parent of the root list expression directly to the currently active list item. Table 1 provides the formal definition of the default outputs for a list expression.

List expressions add four new instructions used to initiate and coordinate list chains:

`ActivateDepth` toggles a list node's activity. It must include a target depth, and like accessing an item in a list by its index, the targeted node switches from inactive to active, thereby connecting its subgraph to the larger list structure. Algorithm 1 illustrates this process. First, all list nodes must receive the instruction. This is ensured by applications passing the instruction to both children, and by the active list node passing it to its CRP. Each list node then compares its current depth to the instruction's target depth. If they match, the node raises its resolve flag, changing its default state. Otherwise, the node sets its resolve flag to 0, preventing multiple subgraphs from being connected at once. The depth targeting property allows programmers to split suspended expressions across multiple distinct separate list chains and provided all the necessary sub-graphs share a common depth. A single instance of this instruction can reconnect all parts of the suspended expression to the graph simultaneously.

**Algorithm 1** Activate Depth

**if** $[EXP]! = [(List)]$ **then**
    $[CLI] \leftarrow [PIB]; [CRI] \leftarrow [PIB]$
**else**
    **if** $[PIB][UNI] == [CRI][UNI]$ **then**
        $[CRI] \leftarrow [PIB]; [RSF] \leftarrow [1]$
    **else**
        $[RSF] \leftarrow [0]$
    **end if**
**end if**

`UpdateDepth` acts similarly to `UpdateChildLeft`, updating a targeted nodes CLP to a new value retrieved from the PEB. The difference being this instruction targets nodes based on depth and can only edit the contents of List nodes. Algorithm 2 shows a pseudocode version of this instruction. If two independent list chains receive this instruction it will cause multiple nodes to update their pointers simultaneously. When sending this instruction ensure that two lists do not receive this instruction and the same new CLP value as this will cause a data collision.

**Algorithm 2** Update Depth

**if** $[PIB][UNI] == [CRI][UNI]$ **then**
    $[CLP] \leftarrow [PEB][CLP]$
**end if**

`AddBottomNode` adds a node to the end of a list. When used together with `RemoveBottomNode` it enables the list to emulate stack-like behavior, where the bottom node is the top of the stack.

| Expression | | Output Values | | | | | |
|---|---|---|---|---|---|---|---|
| | | Parent | | Child Left | | Child Right | |
| | | PIB | PEB | CLI | CLE | CRI | CRE |
| Lists | RSF = 0 | {[CRI]([INS]),Depth+1} | [CRE] | - | - | [PIB] | [PEB] |
| | RSF = 1 | {[CLI]([INS]),Depth+1} | [CLE] | [PIB] | [PEB] | - | - |

Table 1: List Node Default

Algorithm 3 shows this operation. This instruction targets the node with a depth of zero. When a non-targeted node receives this instruction it passes data downwards, eventually reaching the target node. This instruction takes multiple clock cycles to complete. The first clock pulse causes the target node to request a new ID, so it can prepare to link to the new node. During the second clock cycle, the target node updates its CRP to point to the new node. Finally, the target node sends an UpdateExpression instruction to its child, ensuring it adopts the expression characteristics defined on its PEB.

**Algorithm 3** Add Bottom Node

if $[EXP]! = [(Arithmetic)]$ then
    $[CRI] \leftarrow [PIB]; [CLI] \leftarrow [PIB]$;
else
    $[CRI] \leftarrow [PIB]$;
    if $ClockPulses == 0$ then
        $[CRI] \leftarrow [PIB]$;
        if $[CRI][UNI] == 0$ then
            $[CRP] \leftarrow [NewNodeID]$;
        end if
    end if
    if $ClockPulses == 1$ then
        if $[CRI][UNI] == 0$ then
            $[CRI] \leftarrow [UpdateExpression][CRP]$;
        end if
    end if
end if

RemoveBottomNode removes the bottom node from a list. This process is illustrated in the pseudocode shown in Algorithm 4, taking multiple clock pulses to complete and targeting a node of depth one. During the first clock pulse, each node, other than the target, passes this instruction downwards, ensuring that every node receives it. The target begins removal by sending a Nullification instruction to its child. During the second clock pulse, non-targeted nodes continue to pass the instruction and the target sets its CRP to zero, removing its child from the list.

**Algorithm 4** Remove Bottom Node

if $[EXP]! = [(Arithmetic)]$ then
    $[CRI] \leftarrow [PIB]; [CLI] \leftarrow [PIB]$;
else
    if $[CRI][UNI] == 1$ then
        $[CRI] \leftarrow [Nullification]$;
        $[CRp] \leftarrow [0]$;
    end if
end if

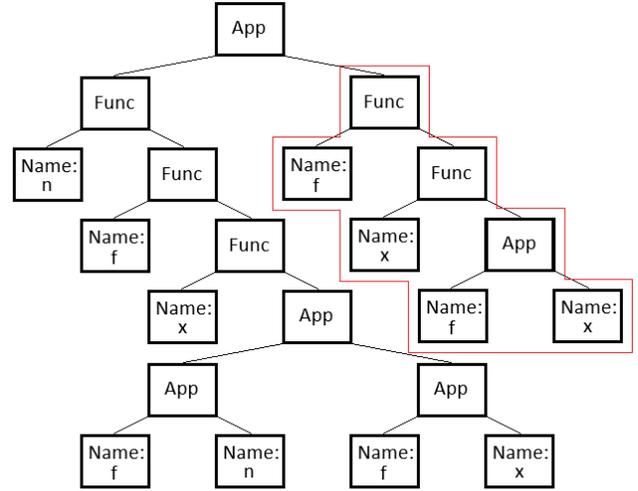

Figure 4: Increment applied to Numeral 1

*Affected Instructions.* We add four new instructions that apply exclusively to list nodes. Several pre-existing instructions also affect these new expressions: Nullification, UpdateExpression, and ReturnExpression can influence lists. However, list expressions do not respond conventionally to ReturnExpression. Instead, they re-send the instruction so that it targets either their left or right child, depending on the lists activity, and then forward the instruction to the appropriate neighbor. This once again links the parent of the root list node directly to the active subgraph, enabling the parent to query that subgraph as though it were directly connected.

## 3 Arithmetic primitives

Church numerals require large space to represent even small numbers. For example, $(\lambda n.\lambda f.\lambda x.f(nfx))(\lambda f.\lambda x.fx)$ represents the increment operation applied to the number 1. When reduced, it becomes the Church numeral for 2. Fig. 4 shows this expression in graph form. The part highlighted corresponds to the number 1. It costs 21 nodes to represent this graph, and many clock ticks to reduce what is effectively 1 + 1. This problem becomes more apparent with larger numbers and more complex arithmetic operations. To overcome it, we will allow name values to be interpreted as two's complement binary numbers by adding expressions capable of performing arithmetic and logical comparisons, allowing programmers to replace these graphs with simpler ones.

The following arithmetic and logical operations are selected as primitive expressions due to frequent use:

Harry Fitchett, Jasmine Ritchie, Charles Fox
School of Engineering and Physical Science,
University of Lincoln, UK

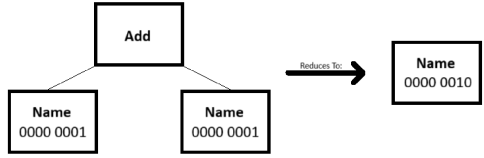

Figure 5: Add expression reducing 1+1

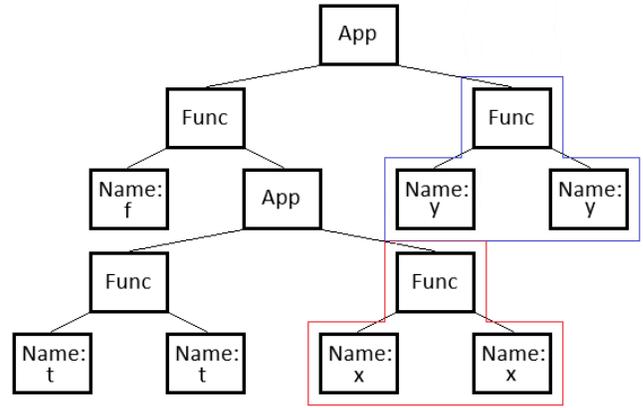

Figure 6: Graph of Decision Point Example

$$\langle expression \rangle ::= \langle name \rangle \mid \langle function \rangle \mid \langle application \rangle$$
$$\mid \langle list \rangle \mid \langle arithmetic \rangle$$
$$\langle arithmetic \rangle ::= \langle add \rangle \mid \langle mult \rangle$$
$$\mid \langle greatzero \rangle \mid \langle lesszero \rangle \mid \langle equalzero \rangle$$
$$\langle add \rangle ::= (\delta + \langle name \rangle \langle name \rangle)$$
$$\langle mult \rangle ::= (\delta \times \langle name \rangle \langle name \rangle)$$
$$\langle greatzero \rangle ::= (\delta > \langle expression \rangle \langle expression \rangle))$$
$$\langle lesszero \rangle ::= (\delta < \langle expression \rangle \langle expression \rangle))$$
$$\langle equalzero \rangle ::= (\delta = \langle expression \rangle \langle expression \rangle))$$

Unlike other expressions, Arithmetic expressions have limited default outputting patterns, actively preventing the reduction of their branches. Table 2 shows these. Instead, an Arithmetic expression, upon receiving resolve flags from specific nodes, performs its arithmetic or logical operation then converts itself into a different expression type.

*Addition and Multiplication* expressions are designed to replace the more complex Successor, Addition, Multiplication, Predecessor, and Subtraction graphs originally proposed by Church for his numeral-based arithmetic. Both expressions assume their child nodes to be Names that are providing values ready for operations. If this condition is not met, the syntax is considered invalid and may cause the graph to behave unpredictably during reduction. When an Addition or Multiplication expression receives two raised resolve flags from its children, it triggers the execution its corresponding arithmetic operation on the values received from both children then upon operation completion nullifying its child nodes and finally transforming itself into a Name expression containing the computed value.

A programmer can use the Addition expression to construct a graph capable of computing arithmetic. Fig. 5 shows 1+1 solved using Addition expressions.

*Comparisons to Zero.* In assembly programming, conditional jump instructions allow a program to jump to a specific line when a condition is met. In a similar way, the GreatZero, LessZero, and EqualZero Arithmetic expressions are here proposed as mechanisms for making conditional decisions. Church defined a set of expressions to perform these operations within his logical system. However, like his arithmetic operations, these approaches are impractical for digital logic implementations due to their large size and high computational cost.

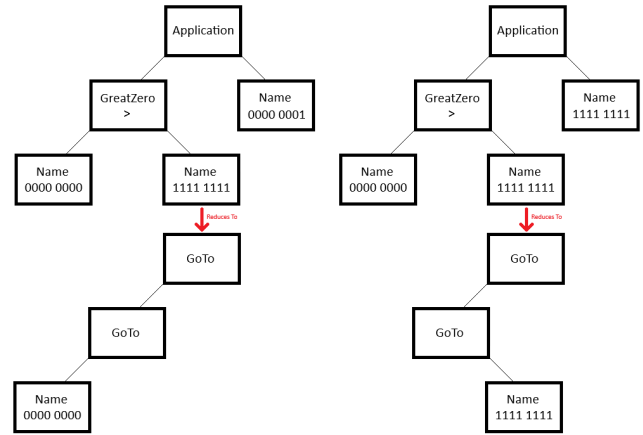

Figure 7: Logical Comparison Arithmetic Reduction

The expression $(\lambda f.(\lambda t.t)(\lambda x.x))(\lambda y.y)$ demonstrates how a decision point can be represented using Church's definitions of true and false. Conceptually, this can be understood as: if true, connect to branch 1; otherwise, connect to branch 2.

This structure is illustrated in Fig. 6, where branch 1 is shown in red and branch 2 is shown in blue.

The GreatZero, LessZero, and EqualZero expressions must connect to three other nodes: a branch 1, a branch 2, and a Name node whose value is compared to zero. The two branches can be connected to as ordinary child nodes, but to support a third connection, similar to a function input, these expressions also require an Ancestor input. When a GreatZero, LessZero, or EqualZero node receives a raised resolve flag from its Ancestor, it compares the Ancestor's value to zero and performs its associated logical comparison, removing its ancestor by using the `ImediateResolution` instruction. If the result of the comparison is true, the node transforms into a GoTo node that points to its left child. If the comparison is false, it instead becomes a GoTo node that points to its right child. This process is represented in Algorithm 5 and demonstrated in Fig. 7.

These arithmetic expressions not only reduce the number of nodes needed to represent Church structures but also introduce a simple form of branch prediction. In the graph shown in Fig. 6, the logical comparison can only proceed once both branches have raised their resolve flags. This means that if one branch is particularly large or complex to reduce, the entire comparison must wait, even if that branch's result will eventually be discarded. In contrast, the GreatZero, LessZero, and EqualZero evaluate lazily compared to Church's strict evaluation method, waiting only for a raised resolve flag from their Ancestor. This allows the node to perform the comparison and transformation even if either branch has not completely reduced, potentially saving computation time by avoiding unnecessary work on irrelevant branches.

| Arithmetic Expression | Output Values | | | | | |
|---|---|---|---|---|---|---|
| | Parent | | Child Left | | Child Right | |
| | PIB | PEB | CLI | CLE | CRI | CRE |
| Add/Mult | - | - | {-,[Raised],-,-} | - | {-,[Raised],-,-} | - |
| GreatZero/LessZero/EqualZero | | | | | | |
|    Reducible | - | - | {-,[Raised],-,-} | - | {-,[Raised],-,-} | - |
|    Irreducible | - | - | {[RSF],[RDF],[CLP],[CRP]} | - | - | - |

Table 2: Arithmetic Node Default

In this current system, while ancestor inputs can be either functions or names, arithmetic expressions that expect an ancestor input assume that the ancestor is a name. Therefore, these expressions rely on their ancestor nodes having reduced to a name by the time a raised resolve flag is sent. If this condition is not met, the graph may reduce unpredictably.

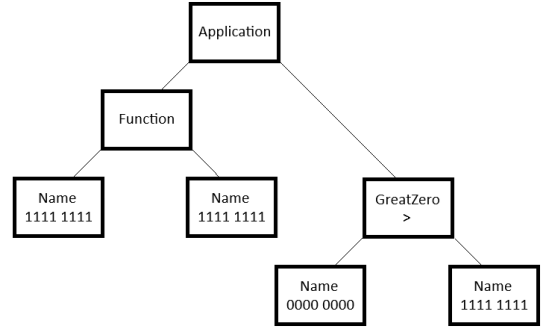

(a) Irreducible Arithmetic Expression as Ancestor Input

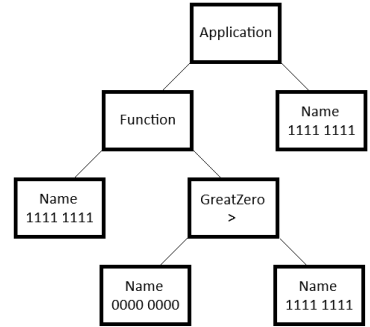

(b) Irreducible Arithmetic Expression as Decsendant Input

Figure 8: Irreducible Arithmetic Expression

---

**Algorithm 5** Zero comparison operations

**if** $[RDF] == 0$ **then**
    **if** $[PEB][CLP], [PEB][CRP] (> or = or <) 0$ **then**
        $[CRP] \leftarrow [(Nullification)]$
    **else**
        $[CLP] \leftarrow [(Nullification)]; [CLP] \leftarrow [CRP]$
    **end if**
    $[PIB] \leftarrow [(ImmediateResolution)]; [EXP] \leftarrow [(GoTo)]$
**else**
    $[RSF] \leftarrow [(Raised)]$
**end if**

---

In some cases, an Arithmetic Expression can be irreducible, meaning it has no ancestor input. When this happens, as in irreducible functions, the arithmetic expression cannot perform its comparison until it becomes reducible. In this specific context a node representing an Arithmetic expression can raise its resolve flag once it receives resolve flags from both of its children, allowing the branch it belongs to continue reducing. The final row of table 2 illustrates the default output patterns for GreatZero, EqualZero, and LessZero Arithmetic Expressions when marked as irreducible. Here, the nodes representing these expressions return their internal contents to their parent through the PEB, acting almost identically to an irreducible function.

*Affected Instructions.* Nodes representing arithmetic expressions are affected by a set of pre-existing instructions. The following instructions influence arithmetic expressions: Nullification, Update Expression, Update Child Left, Update Child Right, and Return Expression. However, when an arithmetic expression determines itself to be irreducible, it may become involved in a graph transformation. To accommodate this, instructions responsible for coordinating function reduction between a function node and its ancestor/descendant inputs must also affect arithmetic expressions that appear within those ancestor or descendant branches.

An irreducible Arithmetic expression could find itself used as an ancestor input for another node. Fig. 8a shows an example of this. In such cases, Arithmetic expressions can be affected by the AncestorTransformation and ImmediateResolution instructions. While any arithmetic expression could theoretically receive these instructions, in practice only irreducible logical comparison expressions will do so. This occurs because arithmetic nodes block reduction until they transform into a different expression type, at which point they raise their resolve flag.

Similarly, an irreducible Arithmetic expression, such as GreaterZero, LessZero, or EqualZero, may exist within a function undergoing reduction, as in Fig. 8b. Here, either branch of the arithmetic expression may contain name nodes that the function attempts to transform. Regardless of which branch is ultimately retained or discarded, both may be subject to transformation. To support this, a node representing an irreducible Arithmetic expression must forward the CompareValue and DescendantTransformation instructions to both of its children. Consequently, not only can arithmetic expressions be affected by these instructions, but the instructions themselves must also be adapted to ensure they are properly forwarded by arithmetic expressions. The additional logic required for this behavior is shown in Fig. 6, which must be included in the definitions of DescendantTransformation and CompareValue.

---

**Algorithm 6** Adaptation to the CompareValue and DescendantTransformation Instruction to Accommodate Arithmetic Expressions

---
**if** $[EXR] == [Arithmetic]$ **then**
$\quad [CLI] \leftarrow [PIB]; [CLE] \leftarrow [PEB]$
$\quad [CRI] \leftarrow [PIB]; [CRE] \leftarrow [PEB]$
**else**
$\quad$ %Instruction_Operates_Like_Normal%
**end if**

## 4 Peripheral Components

Adding new expression types naturally increases the physical size of nodes, as additional logic is required to handle their unique routing and behavior. List and arithmetic expressions demand complex digital logic for operations such as addition, multiplication, and comparison. If every node were equipped with its own adder and multiplier, nodes would grow substantially.

One straightforward approach takes advantage of a common property of most graph structures: the majority of nodes are not arithmetic or list nodes. This observation suggests that only a limited subset of nodes should support these more complex expression types. In this scheme, only specific nodes are capable of representing Arithmetic or List expressions. To accommodate during transformation nodes must be able to request specialized nodes that can represent these specific expression types. This would require the node allocation component to become more sophisticated and instructions involved in the function reduction process would require a redesign. Moreover, as discussed in an earlier chapter, restricting cluster flexibility wastes more space in the form of unused nodes. While this selective-node approach offers a path to reduce complexity and conserve space, it introduces new problems outweighing its benefits.

Another option adopts an approach found in conventional CPU design. Instead of localizing this advanced complicated digital logic to individual nodes. Instead allow groups of nodes, say clusters, to access a shared peripheral component capable of receiving arithmetic/logical requests like a conventional ALU.

Fig. 9 illustrates the circuit diagram of a cluster-wide ALU. It consists of two main components: a standard ALU and a series of shift registers. The ALU performs arithmetic and logical operations on two operands based on a provided operator, while the shift registers are capable of storing a unique node ID, two operands, and an opcode. These registers function as a stack, allowing the circuit to queue and manage multiple ALU requests from different nodes.

When a node needs to perform an operation, it sends its unique ID, the two operand values, the desired operation, and a one-bit signal indicating a request. From the perspective of the circuit, the node's data (the ID, operands, and operation) are received through the D input, while the one-bit request signal is received through the s0 selector bit. When this signal is active, each shift register shifts its contents down by one position, and the top register stores the new values arriving through the D input.

On the next clock pulse, if no new requests have been received, the ALU begins processing the operation stored in the topmost register. Once the ALU completes, it sends a high signal through its status output, triggering two actions. First, the result (ALU Q output) is sent to the node whose Unique Node ID matches the value stored in the ID field. Second, each shift register shifts its contents again, this time moving the next pending request upwards to the top of the stack for processing.

It might make sense to include a second, or more, ALU components to handle computation requests more efficiently, at the cost of increasing the size and complexity of work clusters. This trade-off would depend on task-specific requirements.

Determining the optimal number of ALUs per peripheral cluster is a balancing act that depends heavily on implementation. We recommend limiting the number of ALUs to one ALU per cluster. Instead of adding more ALUs, it's often better to reduce the number of nodes in each cluster, which limits how many nodes can send requests to a single ALU. This approach keeps the cluster's overall size manageable, since cluster size mainly depends on node count and peripheral component complexity, while limiting the number of nodes each ALU is required to service.

Arithmetic expressions use the ALU in the usual way: when a triggering condition is met, a request is sent to the ALU, which processes it and performs the necessary transformation based on the result.

Lists use arithmetic differently, using arithmetic operations to maintain a continuous depth value across a chain of list nodes. As this signal is continuous, each List expression requires a local, adder to calculate and preserve this property. While not ideal, these adders are the most light weight arithmetic logic circuit available capable only capable of incrementing a value, so not increasing node bulk.

## 5 Results

As in prior work [20], the system was evaluated using Logisim Evolution, with implementations and solutions available as open source.[1] [2] Table 3 summarises a set of representative expressions that can be reduced using the simulation. Each expression is provided as a test bench in the accompanying repository, enabling full

---
[1] https://github.com/LAMB-TARK/Extending-CPU-less-parallel-execution-of-Lambda-calculus-in-digital-logic-towards-practicality
[2] a video demo is provided at https://www.youtube.com/watch?v=1_H69wHMYO8

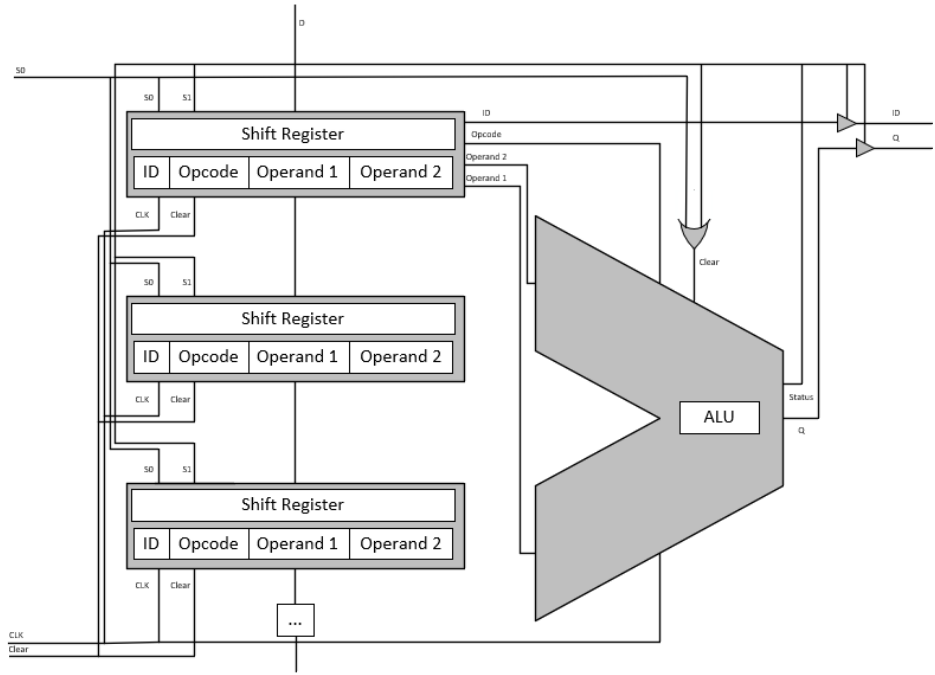

Figure 9: Cluster ALU Circuit Diagram

| Index | Expression | n nodes | ticks | depth | n alt nodes | result | correct? |
|---|---|---|---|---|---|---|---|
| 1 | $(\delta + 1.1)$ | 3 | 16 | N/A | 31 | 2 | Success |
| 2 | $(\delta + (\delta + 1.1).(\delta + 1.1))$ | 5 | 48 | N/A | 91 | 4 | Success |
| 3 | $(\delta * 3.3)$ | 3 | 16 | N/A | 45 | 9 | Success |
| 4 | $(\delta < a.b) - 1$ | 5 | 16 | N/A | 23 | a | Success |
| 5 | $(\delta < a.b)1$ | 5 | 16 | N/A | 23 | b | Success |
| 6 | $(\delta == a.b)0$ | 5 | 16 | N/A | 23 | a | Success |
| 7 | $(\delta == a.b)1$ | 5 | 16 | N/A | 23 | b | Success |
| 8 | $(\delta > a.b)1$ | 5 | 16 | N/A | 23 | a | Success |
| 9 | $(\delta > a.b)1$ | 5 | 16 | N/A | 23 | b | Success |
| 10 | $(\lambda f.f)(\gamma a.(\gamma b.\emptyset))$ | 8 | 17 | 0 | 57 | a | Success |
| 10 | $(\lambda f.f)(\gamma a.(\gamma b.\emptyset))$ | 8 | 17 | 1 | 57 | b | Success |
| 11 | $(\lambda f.f)(\gamma(\delta + 3.3).(\gamma(\delta * 3.3).\emptyset))$ | 12 | 24 | 0 | 145 | a | Success |
| 11 | $(\lambda f.f)(\gamma(\delta + 3.3).(\gamma(\delta * 3.3).\emptyset))$ | 12 | 24 | 1 | 145 | e | Success |
| 12 | $(\gamma(\lambda f.f).(\gamma(\lambda f.e).\emptyset))a$ | 10 | 6 | 0 | 31 | a | Success |
| 12 | $(\gamma(\lambda f.f).(\gamma(\lambda f.e).\emptyset))a$ | 10 | 17 | 1 | 31 | e | Success |

Table 3: Test Expression and Results

reproducibility. To quantify the impact of the new constructs, we record several metrics for each test expression:

*Node Count and Clock Pulse.* report the number of nodes and clock pulses required to complete each reduction.

*Depth.* expressions that make use of lists require an activated depth; this value is recorded in the Depth column when applicable.

*Alt Node.* provides the theoretical node count that would be required if the same expression were encoded purely in raw $\lambda$-calculus, without lists or arithmetic expressions.

*Result and Reduction Success.* indicate the final reduced expression and whether the reduction completed successfully.

A limitation of the simulator is that it cannot reduce the equivalent raw $\lambda$ versions of these expressions. Every alternative formulation shown in the table exceeds the simulator's node limit of 16; consequently, no alternative clock counts are available for comparison.

Each expression shows a significant reduction in nodes used and presumably clock pulses required. For example, expressions 1 can be represented using only three nodes, and, unlike Church numerals, their node count does not grow with the magnitude of the numbers. 127 + 127 (the largest addition a single node can process in this system) still requires only three nodes and completes in the same amount of time. The same operation using Church numerals would require 1,041 nodes.

## 6 Discussion

The results show that it is possible and practical to extend the parallel $\lambda$-calculus to digital logic compiler of [20] with list and arithmetic primitives, resulting in substantial reductions in execution time and node usage in a number of test cases.

Arithmetic expressions have allowed clusters to emulate behavior that a conventional CPU would normally delegate to an ALU in an efficient and timely manner. At present, the expression supports a minimal arithmetic operation set expected of an ALU. Yet, ALUs typically also provide support for logical operations. To further develop this system, a corresponding logical expression should be designed to support operations such as logical shifts and masking.

List expressions not only allow programmers easy access to lists but add additional functionality. When implemented in a practical capacity, clusters will have a finite node limit. Therefore, something similar to an operating system must be able to schedule and suspend branches to prevent clusters exceeding this limit. Lists, capable of disconnecting and suspending sections of graphs, enable this. Using lists, an operating system could prioritize branches deemed more urgent by suspending active branches when more urgent tasks arrive. Likewise, lists can be used to suspend all reduction by activating an invalid list depth, allowing for on-the-fly memory editing, providing and equivalent to how a CPU may handle interrupts.

Despite the practical benefits lists provide our system, there is still room for further refinement. To minimize node overhead, the current implementation re-purposes the PIB to transmit depth information instead of instruction data. This can create ambiguity, causing non-list nodes to interpret depth values as instructions. The test bench 12 expression reduces correctly, but produces errors during the read back. In this expression a list passes a depth value of two to its ancestor. However, the node receiving this value is not a list but instead a name which equates the depth value to its corresponding instruction, ReturnExpression. This means as the cluster attempts to read its contents conflicting ReturnExpression instructions are received causing multiple nodes to attempt to return their contents creating a data collision. Mitigation might include expanding the bus width to carry a dedicated depth field.

The current implementation uses 16 nodes per cluster to allow unique IDs to fit conveniently into 4 bits. But is 16 an efficient or desirable number? [20] has many expressions it cannot compute due to exceeding a depth limit. So arguably the more nodes, the more expressions a cluster will be able to compute. However, more nodes means more-node-to-node connections, increasing the size of clusters exponentially. The inclusion of cluster-wide peripheral components – such as the Cluster ALU discussed here – complicates this further. Before any large-scale implementation can be designed, a clear evaluation metric must be developed to guide the optimal number of nodes per cluster.

Additionally, we introduced several new expression types. We argued for their inclusion due to their common appearance in low-level programming languages and, through their inclusion, we created test benches which by contrast to the model of [20] require significantly less time and nodes to reduce. This raises the question of whether additional expression types could or should be added, and how cost and benefit of each addition could be assessed. This is a roughly analogous question to the RISC vs CISC debate in CPU design, in which adding more functionality is traded against the costs of silicon resources. However in CPUs, such costs are typically only incurred once, in a single location, while in our tree structures any small size increase to nodes is instantiated in all the nodes so may have larger size impact.

An equally integral role in the success of any architecture is a compiler. and while many functional compilers exist, they mostly focus on translating functional languages to serial machine code. Our system will require compilers capable of doing the opposite, Translating serial high level languages into parallel functional low-level code. A related system [21] compiles an ad-hoc functional subset of Python all the way to parallel digital logic, demonstrating what a more usable, normal-looking language front-end could look like. The compilation methods presented in [21] could likely be co-opted into producing parallel functional low-level code used by our system.

Despite the widespread theoretical adoption of $\lambda$−calculus, it was never intended as a practical programming language. Many language designers notice that when treated as one, $\lambda$−calculus exhibits a number of desirable qualities. However, when implemented, its simplistic design produces a number of shortcomings. So languages tend to evolve away from raw $\lambda$-calculus, adding and modifying expressions, which typically sacrifices theoretical integrity but accommodates better memory usage, processing time and user access. Abandoning our adherence to strict $\lambda$-calculus might similarly benefit our new architecture. Increasing the number of parent and child connections a singular node can have may reduce the quantity of application nodes required, thereby saving memory. Introducing expressions with more complex but specific functionality would reduce processing time and moving to a more conventional language syntax would allow for better user access.